\documentclass[]{spie}  %>>> use for US letter paper

\usepackage{amsmath,amsfonts,amssymb}
\usepackage{graphicx}
\usepackage[colorlinks=true, allcolors=blue]{hyperref}

\title{Bright Star Subtraction Pipeline for LSST: Phase 1 report}

\author[a, b, c]{Amir E. Bazkiaei}
\author[d]{Lee S. Kelvin}
\author[c, e]{Sarah Brough}
\author[a, b, c]{Simon J. O'Toole}
\author[f]{Aaron~Watkins}
\author[d]{Morgan A. Schmitz}
\affil[a]{Australian Astronomical Optics, Macquarie University, North Ryde, NSW 2113, Australia}
\affil[b]{Astrophysics and Space Technologies Research Centre, Macquarie University, Macquarie Park, NSW 2109, Australia}
\affil[c]{Australian Research Council Centre of Excellence for All Sky Astrophysics in 3 Dimensions (ASTRO 3D), Australia}
\affil[d]{Department of Astrophysical Sciences, Princeton University, Princeton, NJ 08544, USA}
\affil[e]{School of Physics, University of New South Wales, Sydney, NSW 2052, Australia}
\affil[f]{Centre for Astrophysics Research, School of Physics, University of Hertfordshire, Hatfield, UK}

\authorinfo{Further author information: (Send correspondence to A.E.B.)\\A.E.B.: E-mail: amir.ebadati-bazkiaei@mq.edu.au}

\begin{document} 
\maketitle

\begin{abstract}
We present the phase 1 report of the Bright Star Subtraction (BSS) pipeline for the Vera C. Rubin Observatory's Legacy Survey of Space and Time (LSST).
This pipeline is designed to create an extended PSF model by utilizing observed stars, followed by subtracting this model from the bright stars present in LSST data.
Running the pipeline on Hyper Suprime-Cam (HSC) data shows a correlation between the shape of the extended PSF model and the position of the detector within the camera's focal plane. Specifically, detectors positioned closer to the focal plane's edge exhibit reduced circular symmetry in the extended PSF model.
To mitigate this effect, we present an algorithm that enables users to account for the location dependency of the model.
Our analysis also indicates that the choice of normalization annulus is crucial for modeling the extended PSF.
Smaller annuli can exclude stars due to overlap with saturated regions, while larger annuli may compromise data quality because of lower signal-to-noise ratios.
This makes finding the optimal annulus size a challenging but essential task for the BSS pipeline.
Applying the BSS pipeline to HSC exposures allows for the subtraction of, on average, 100 to 700 stars brighter than 12th magnitude measured in g-band across a full exposure, with a full HSC exposure comprising $\approx$100 detectors.
\end{abstract}

\keywords{Vera C. Rubin Observatory, Extended PSF Model, Low Surface Brightness, Galaxies, Bright Star Subtraction}

\section{INTRODUCTION}
\label{sec:intro}

The Vera C. Rubin Observatory's Legacy Survey of Space and Time (LSST) will reach exceptional deep surface brightness limits in its imaging of the southern sky \citenum{2019ApJ...873..111I}, offering a remarkable opportunity to study Low Surface Brightness (LSB) structures around galaxies [e.g., \citenum{2020arXiv200111067B}].
In these deep images, bright stars can cause an over-estimation of the sky background if they are not accurately modeled and subtracted to sufficiently large radii before background measurements are taken \citenum{2023MNRAS.520.2484K, 2024MNRAS.528.4289W}, which will lead to over-subtraction of the sky background.
This over-subtraction removes LSB structures in the deep images expected from LSST \citenum{1991ApJ...369...46U, 2009PASP..121.1267S, 2020MNRAS.491.5317I, 2021ApJ...910...45M, 2021MNRAS.502.2419F, 2022PASJ...74..247A, 2023MNRAS.518.1195M}.

We are developing the Bright Star Subtraction (BSS) pipeline for LSST \citenum{2024arXiv240404802B} to minimize sky over-estimation.
By accurately modeling the stellar profiles of bright stars over several hundred arcseconds, the BSS pipeline will produce an extended Point Spread Function (PSF) model, which will then be used to subtract bright stars from deep images before estimating the sky background.
An important advantage of this pipeline is its ability to increase the effective area of the LSST by reducing the extent of the regions masked around bright stars.
For instance, star masks cover approximately 20$\%$ of the survey area in the HSC Subaru Strategic Program (SSP) survey \citenum{2018PASJ...70S...4A}.
Moreover, extended PSF models can help remove scattered light from bright sources near faint features like intragroup and intracluster light e.g. \citenum{2023arXiv230916244G}.

In this study, we utilize data from the HSC SSP survey to evaluate the bright star subtraction pipeline.
HSC comprises 104 science detectors, with a field of view of 1.5$^\circ$ diameter \citenum{2018PASJ...70S...1M}.
The Release Candidate 2 (RC2) dataset \citenum{DMTN-091}, which is a subset of HSC data, covers approximately 5 square degrees.
This subset consists of about 150 visits which are observed with five broad band filters (\textit{g, r, i, z, y}).
Produced by the LSST Data Management (DM) team, RC2 provides an adequately large dataset for testing ongoing algorithmic changes to the LSST Science Pipelines, making it ideal for evaluating the Bright Star Subtraction pipeline.

\section{Bright Star Subtraction Pipeline Tasks}

\begin{figure} [ht]
   \begin{center}
   \includegraphics[width=\textwidth]{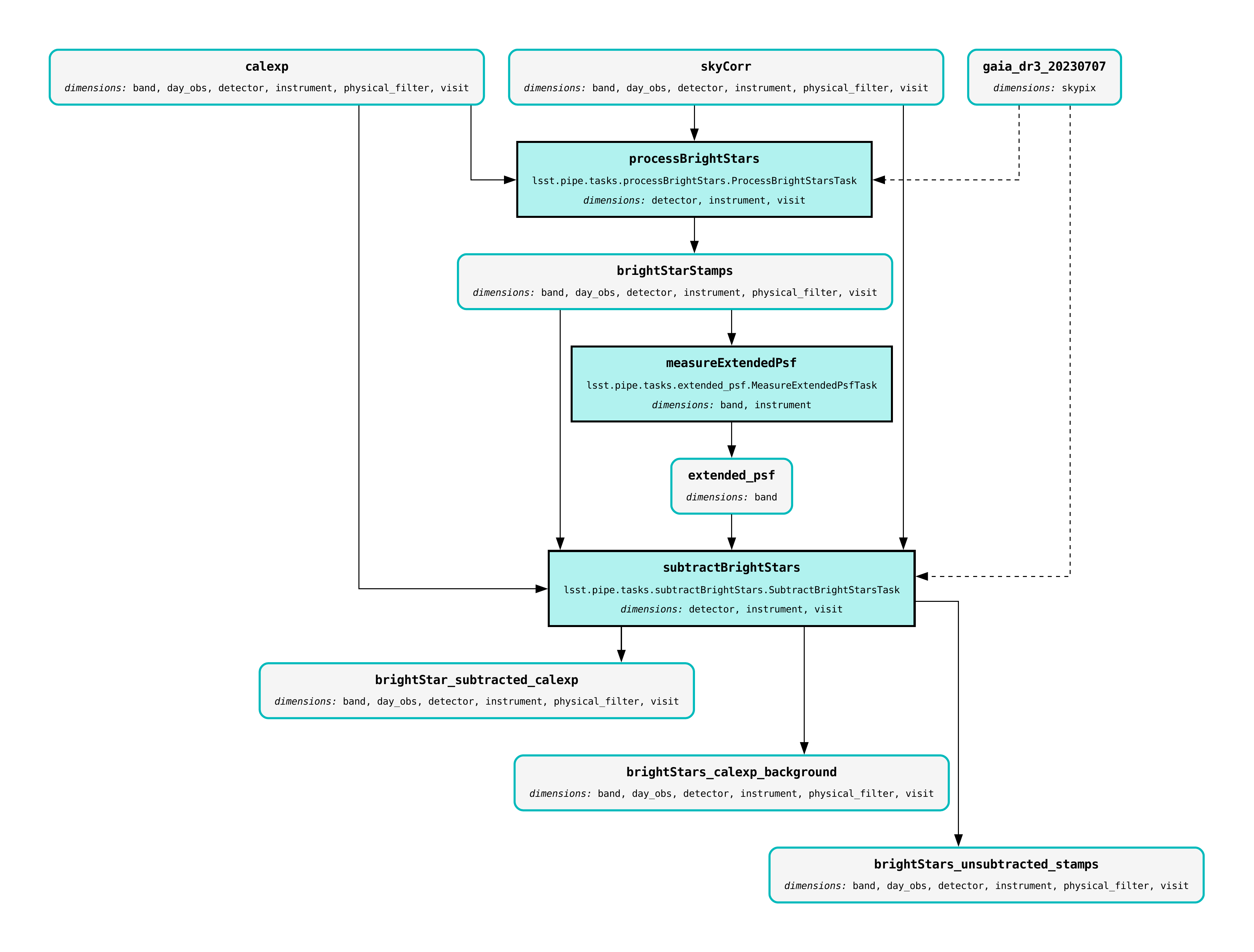}
   \end{center}
   \caption[example] 
   { \label{fig:bss_pipeline} 
A visual representation of the three primary tasks in the LSST Science Pipelines which constitute the bright star subtraction pipeline.
Tasks (shaded teal boxes) and both input and output datasets (turquoise outlined boxes) are shown for reference.}
   \end{figure}

At the end of phase one, the BSS pipeline consists of three main tasks to generate an extended PSF model and subtract the scaled model from bright stars in calibrated exposures.
Figure \ref{fig:bss_pipeline} illustrates the tasks along with their inputs and outputs.
Each of these tasks is a Python class written in a different module.
The modules are located in the \texttt{pipe$\_$tasks} repository\footnote{\url{https://github.com/lsst/pipe_tasks}}, which is the LSST repository containing many of the task classes that drive the LSST Science Pipelines\footnote{\url{https://pipelines.lsst.io/}} \citenum{2018PASJ...70S...5B, 2019ASPC..523..521B}.
The three tasks are:

\begin{itemize}

    \item \texttt{ProcessBrightStarsTask}, in \texttt{processBrightStars.py},
    \item \texttt{MeasureExtendedPsfTask}, in \texttt{extended$\_$psf.py}, and
    \item \texttt{SubtractBrightStarsTask}, in \texttt{subtractBrightStars.py}. 

\end{itemize}

Each of the BSS tasks is a highly configurable Python class.
The net result of this pipeline is the generation of a 2D extended PSF model (i.e. out to several hundred arcseconds) and the subtraction of the scaled model from bright stars in calibrated exposures.
The sections below discuss each task in more detail.
The noted tasks make use of the algorithms provided in \texttt{brightStarStamps.py}.
This module lives in the \texttt{meas$\_$algorithms}\footnote{\url{https://github.com/lsst/meas_algorithms}} package.

\subsection{\texttt{ProcessBrightStarsTask}}

This task produces normalized cutouts (`stamps') by using calibrated exposures and the Gaia reference catalog \citenum{2016A&A...595A...1G, 2023A&A...674A...1G, 2023A&A...674A..32B} as inputs.
By default, per-detector background solutions are removed from the calibrated exposure prior to processing and the full focal plane sky correction (\texttt{skyCorr}) background models are applied.
The data butler \citenum{2022SPIE12189E..11J} is used to identify stars that are brighter than a user-defined magnitude threshold.
A cutout centered on each identified star, with dimensions specified by the user, is then created.
The stamps are then normalized by utilizing the annular flux, which refers to the flux within a user-defined annulus centered on the star.
Optionally, a minimum valid pixel fraction within the annulus can be specified, ensuring only successful annuli are retained.
For a given calibrated exposure, the task produces a single FITS file containing the normalized stamps and their associated mask planes. 
Information on the stellar ID, magnitude and annular flux are persisted in the FITS header.

\begin{figure} [ht]
   \begin{center}
   \includegraphics[width=\textwidth]{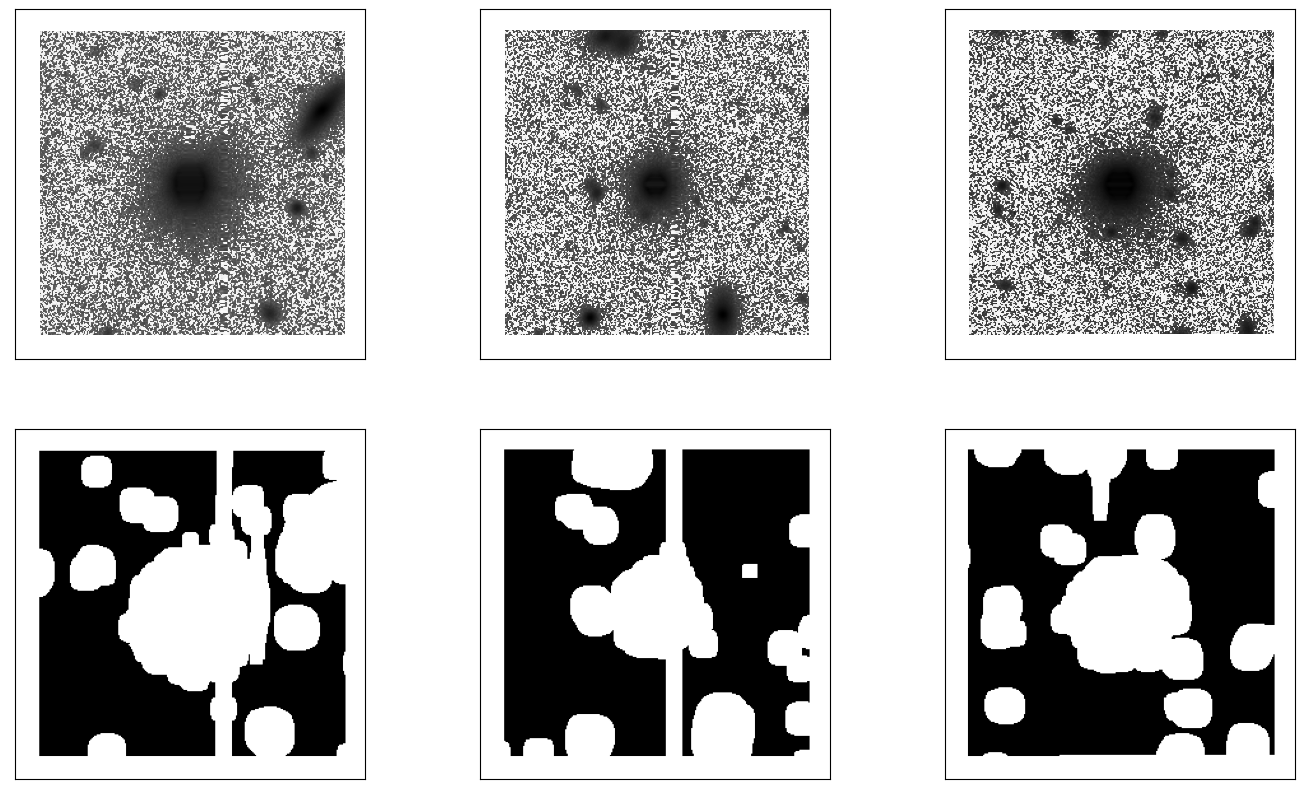}
   \end{center}
   \caption[example] 
   { \label{fig:stamps_default} 
Three examples of normalized stamps produced by \texttt{ProcessBrightStarsTask} are shown in the top row, with their corresponding mask planes in the bottom row. These stamps were generated using the default configuration, which means they are 250 pixels on side.}
   \end{figure}

This task has a number configuration parameters which the user can set.
Some of the important configuration parameters and their default values are: \texttt{magLimit} (magnitude limit), with a default value of 18th magnitude; \texttt{stampSize} (stamp dimensions), with a default value of (250, 250) in pixels; \texttt{annularFluxRadii} (normalization annulus inner and outer radii), with a default value of (70, 80) in pixels; and \texttt{minValidAnnulusFraction} (limit for the minimum fraction of valid pixels), with a default value of zero.
We use an annulus for normalising stamps because the central regions of stars are saturated and useless.
Figure \ref{fig:stamps_default} shows examples of stamps produced by running the \texttt{ProcessBrightStarsTask} with the default configuration.
It is important to acknowledge that the default configurations are established to facilitate programming testing.
Consequently, refining the configuration parameters becomes imperative for scientific testing purposes.

To generate the extended PSF models, we adjust the default configurations and explore the results.
Specifically, we set the magnitude limit to 12, the stamp dimensions to (1000, 1000) pixels, which corresponds to about 170 arcseconds on a side, and tested three different normalization annulus radii (inner, outer): (100, 140), (140, 180), and (180, 220) pixels.
The minimum valid annulus fraction is maintained at zero.
These adjustments allow for construction of a more comprehensive set of star stamps suitable for extended PSF generation.

   \begin{figure} [ht]
   \begin{center}
   % \begin{tabular}{c} %% tabular useful for creating an array of images 
   \includegraphics[width=\textwidth]{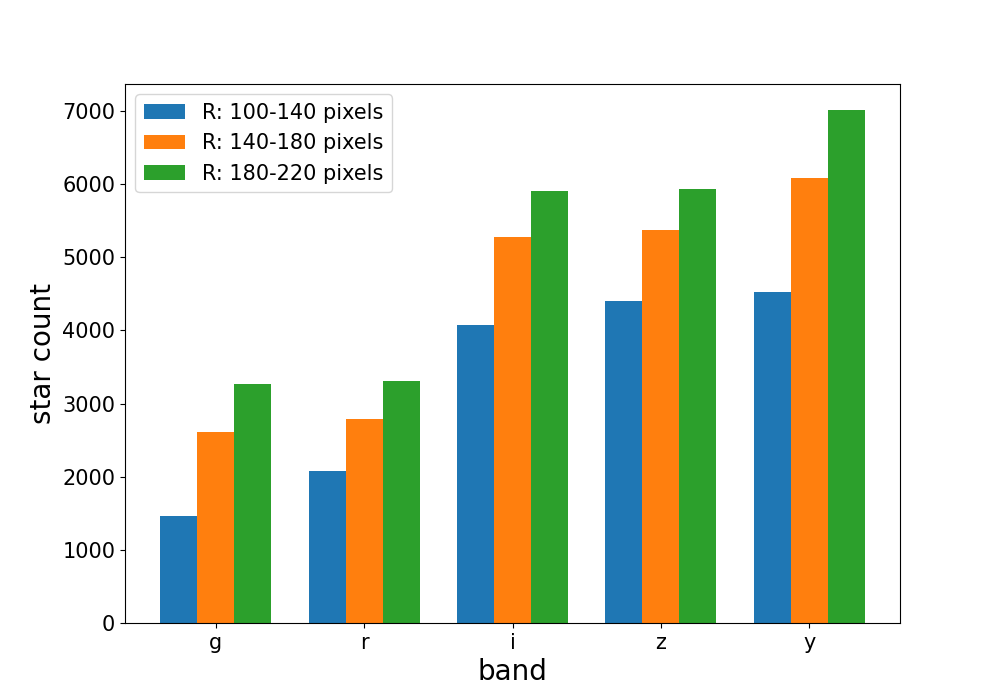}
   % \end{tabular}
   \end{center}
   \caption[example] 
   { \label{fig:starCounts} 
The number of stars with a valid annular flux in different bands and for different annular radii.
The number of stars increases for larger annuli. 
This is due to the fact that saturated central regions of stars are less likely to overlap with the annulus in such cases, allowing for a greater number of valid pixels to be counted.}
   \end{figure}

The size of the normalization annulus significantly impacts the set of stars included in the process.
Figure \ref{fig:starCounts} illustrates the number of stars counted in each annulus for three colors and different bands.
It shows that the number of stars increases with larger normalization annuli.
Conversely, smaller normalization annuli result in more stars being excluded due to a lack of sufficient viable pixels.

One of the reasons smaller annuli exclude more stars is that they are more likely to overlap with the saturated central regions of the stars.
Since these saturated areas are masked, they do not contribute valid pixels to the annular flux measurement, resulting in fewer stars meeting the criteria.
Larger annuli cover a greater area, increasing the probability of finding valid pixels, albeit at lower flux levels.
This expanded area is why larger annuli are more likely to include a broader range of stars in the process.

  \begin{figure} [ht]
   \begin{center}
   \includegraphics[width=\textwidth]{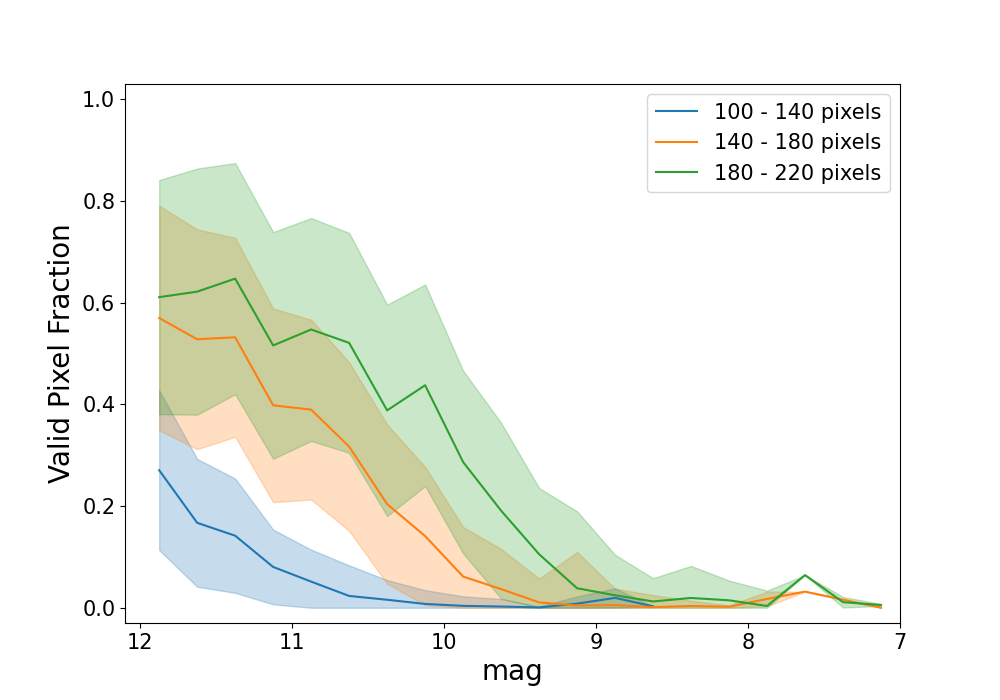}
   \end{center}
   \caption[example] 
   { \label{fig:i_valid_pixel_fraction}
The fraction of valid pixels within the normalization annulus for three different annulus sizes in HSC i-band observations.
Generally, larger annuli contain a greater fraction of valid pixels.
Additionally, the fraction of valid pixels tends to decrease for brighter stars.}
   \end{figure}

Figure \ref{fig:i_valid_pixel_fraction} compares the fraction of valid pixels within the normalization annulus for HSC i-band observations with different annulus sizes.
The largest annular region (in green) has the highest valid pixel fraction, while smaller annuli show lower fractions of valid pixels.
This trend is also influenced by star brightness; brighter stars tend to have a lower fraction of valid pixels due to larger saturated regions overlapping with the annulus.

\begin{figure} [ht]
   \begin{center}
   \includegraphics[width=\textwidth]{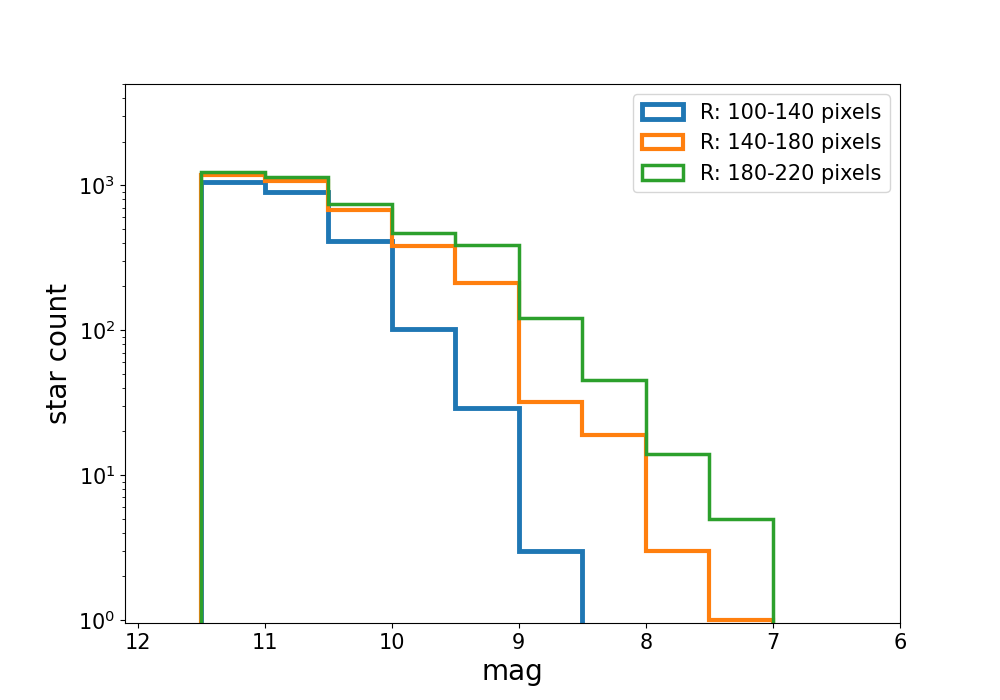}
   \end{center}
   \caption[example] 
   { \label{fig:iband3_hist} 
The number of stars with a positive annular flux, binned according to star magnitude and and for different normalization annulus radii.
The number of stars decreases for for brighter stars regardless of the annulus radius.
However, larger normalization annuli include more bright stars compared to smaller normalization annuli.}
   \end{figure}

The overlap between the saturated central regions and the normalization annulus often leads to excluding brighter stars from the process, effectively setting a lower limit on the brightness of stars that can be included.
Figure \ref{fig:iband3_hist} shows the histogram of stars with positive annular flux for different annulus sizes.
As the plot shows, the number of stars increases with decreasing brightness until the drop to zero.
Moreover, the number of brighter stars decreases with smaller annuli.
Larger annuli, on the other hand, are more inclusive of brighter stars as they reduce the overlap with the saturated region.

\begin{figure} [ht]
   \begin{center}
   \includegraphics[width=\textwidth]{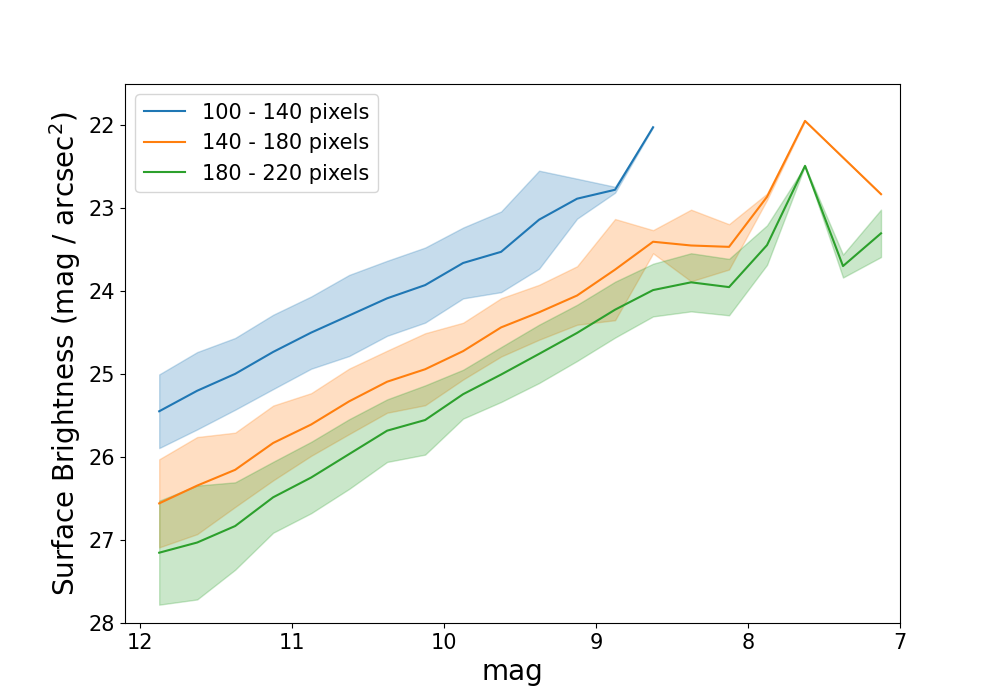}
   \end{center}
   \caption[example] 
   { \label{fig:i_SB} 
Annular surface brightness of stars plotted as a function of the Gaia magnitude of stars (mag) for three different normalization annuli.
Larger annuli tend to show fainter annular surface brightnesses, as these typically encompass lower signal-to-noise ratio data.
The annular surface brightness also decreases with decreasing star brightness for all annuli radius values presented here, as expected.}
   \end{figure}

Choosing larger normalization annuli clearly increases the number of stars included in the process, but this has the potential to impact the quality of the resultant extended PSF models.
We explore this in Figure \ref{fig:i_SB} which shows the annular surface brightness of stars against their Gaia magnitudes for stamps with different normalization annuli.
This figure shows that larger annuli tend to result in fainter annular surface brightnesses, suggesting that larger annuli use data with lower signal-to-noise ratios as compared to smaller ones.
This underscores the importance of considering the size of the normalization annulus to ensure a good balance between star inclusion and data quality.

Depending on the size of the normalization annulus and the observed band, this analysis indicates that the BSS pipeline subtracts, on average, from one to seven stars from the image of each detector.
Given that there are 104 detectors in HSC, this results in the order of between 100 and 700 stars subtracted from a full HSC exposure; a value which will vary depending on the selected annulus size and band, and star magnitude subtraction threshold.

In summary, selecting the appropriate size of the normalization annulus is critical for the success of the bright star subtraction process.
Larger annuli increase the number of stars that can be included in the process, but they may compromise the quality of the data due to lower signal-to-noise ratios.
Smaller annuli tend to exclude more stars, particularly brighter ones, due to the overlap with saturated central regions.
Thus, careful consideration must be given to annulus size when configuring the bright star subtraction pipeline.
Ultimately, achieving a balance between star inclusion and data quality is essential for generating accurate extended PSF models and ensuring robust star subtraction.

\subsection{\texttt{MeasureExtendedPsfTask}}

This task stacks normalized star stamps to generate an extended PSF model.
Various statistical methods for stacking are available, including median, mean, and sigma-clipped mean (default) with a user-defined outlier rejection sigma (the default sigma-clip value is $\sigma=4$).

\begin{figure} [ht]
   \begin{center}
   \includegraphics[width=\textwidth]{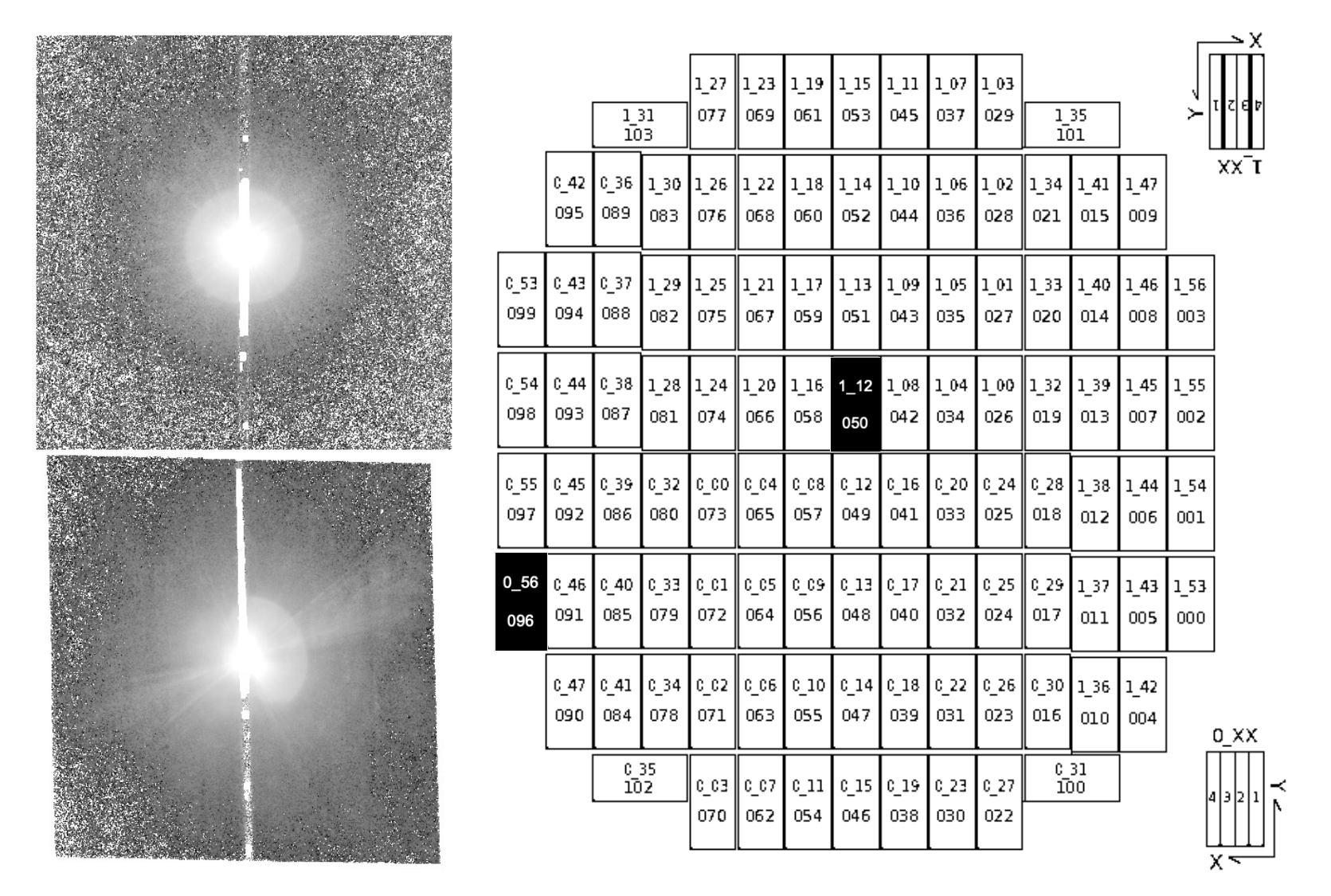}
   \end{center}
   \caption[example] 
   { \label{fig:hsc_focal_plane} 
The extended PSF models for a central detector (050) and a peripheral detector (096) are displayed in the top left and bottom left panels, respectively. The right panel presents an overview of the detectors on the HSC focal plane, adapted from the Subaru Telescope website\addtocounter{footnote}{-1}\protect\footnotemark. Each detector is labeled with its ID. Detectors whose models are shown on the left are highlighted in black in the right panel. Notably, the model for the central detector (050) exhibits circular symmetry, contrasting with the model for the peripheral detector (096). This difference illustrates the model’s dependency on the detector’s position on the focal plane.}
   \end{figure}

The extended PSF models developed for individual HSC detectors exhibit a significant dependency on the detector's position within the focal plane \citenum{2024arXiv240404802B} as illustrated in Figure \ref{fig:hsc_focal_plane}.
The asymmetry of the extended PSF model increases notably as the detector moves farther away from the center of the focal plane.

\footnotetext{\url{https://www.subarutelescope.org/Observing/Instruments/HSC/hsc_ccd_anomaly.html}}

In order to mitigate the impact of detector location dependency on the extended PSF model, users have the option to define detector ``regions" on the focal plane.
Each region can encompass one or more detectors, thereby potentially enhancing signal-to-noise ratios while sacrificing certain positional information.
Stamps from all detectors within a region are aggregated to construct a 2D extended PSF model specific to that region.
These regional extended PSF models are subsequently employed to remove bright stars from calibrated exposures within the corresponding region.

If the choice is made to generate an extended PSF model for each detector, as with HSC observations, the number of stars available to create the model is significantly reduced.
In this case, the star count for each detector is approximately one percent of the count when generating a focal-plane-level model.
This substantial reduction in the number of stars can lead to less robust extended PSF models and increase the risk of artifacts due to limited data.
This can of course be mitigated by using more data where available, for example, a greater number of HSC visits.

It is important to note that the Bright Star Subtraction (BSS) pipeline and the \texttt{MeasureExtendedPsfTask} are not designed to model the core of the Point Spread Function (PSF).
The primary focus of this task is on modeling and correcting for the extended PSF, with a focus on the outer regions of the light distribution around bright stars.
The core of the PSF, which typically involves a more compact light distribution, is modeled and handled by \texttt{MeasurePsfTask} of the LSST Science Pipelines.
This task is located inside the \texttt{measurePsf.py} module in \texttt{pipe$\_$tasks} \citenum{2018PASJ...70S...5B, 2019ASPC..523..521B}.

\subsection{\texttt{SubtractBrightStarsTask}}

The \texttt{SubtractBrightStarsTask} operates on the output of the \texttt{MeasureExtendedPsfTask}, which is the generated 2D extended PSF model.
This task involves scaling the PSF model to align with the luminosity level of a star, followed by the subtraction of the scaled model from the corresponding calibrated exposure.
The identification of stars for subtraction involves querying the data repository \citenum{2022SPIE12189E..11J} for relevant metadata associated with each star within the calibrated exposure.
Leveraging the Gaia reference catalog, stars within a user-defined luminosity threshold are identified for subtraction.
Upon completion of the subtraction process for all identified stars, a subtracted calibrated exposure is produced.
Each bright-star-subtracted calibrated exposure is archived in the data butler as a data product, available for retrieval upon request in the form of  a FITS file.

There are two types of extended PSF scaling available: the \texttt{annularFlux} and the \texttt{leastSquare} algorithms, with the latter being the default.
The \texttt{annularFlux} algorithm operates by multiplying the extended PSF model with the star's annular flux.
On the other hand, the \texttt{leastSquare} algorithm works by determining the least square scaling factor and then multiplying the model by this factor.

Subtraction may not occur during the first pass for some stars, typically due to one of two factors.
Firstly, these stars could have been marked as rejected when the normalized stamps were initially generated using the \texttt{ProcessBrightStarsTask}.
Second, the magnitude threshold for star subtraction might be greater than the original magnitude limit selected for creating the normalized star stamps.
To address these issues, the pipeline incorporates adjusted methods specifically designed to handle these situations and subtract these stars.

Since rejected stars lack a valid flux within the normalization annulus, this task progressively tests larger annuli to identify a feasible flux.
The first annulus in this step has an inner radius that matches the outer radius of the normalization annulus defined by the user in \texttt{ProcessBrightStarsTask}.
The annulus width remains constant.
Should no flux be found within this expanded annulus, the process repeats, extending out to a larger annulus.
This search for a valid flux measurement persists as long as the outer radius of the annulus is within the edge of the stamp.
If no viable flux value is found, neither modelin nor subtraction of the star occurs.
For stars fainter than the originally chosen magnitude limit for generating normalized star stamps, the task follows a similar process, commencing from the normalization annulus.
The task generates a single output containing stamps of unsubtracted stars along with each subtracted calibrated exposure.

\begin{figure} [ht]
   \begin{center}
   \includegraphics[width=\textwidth]{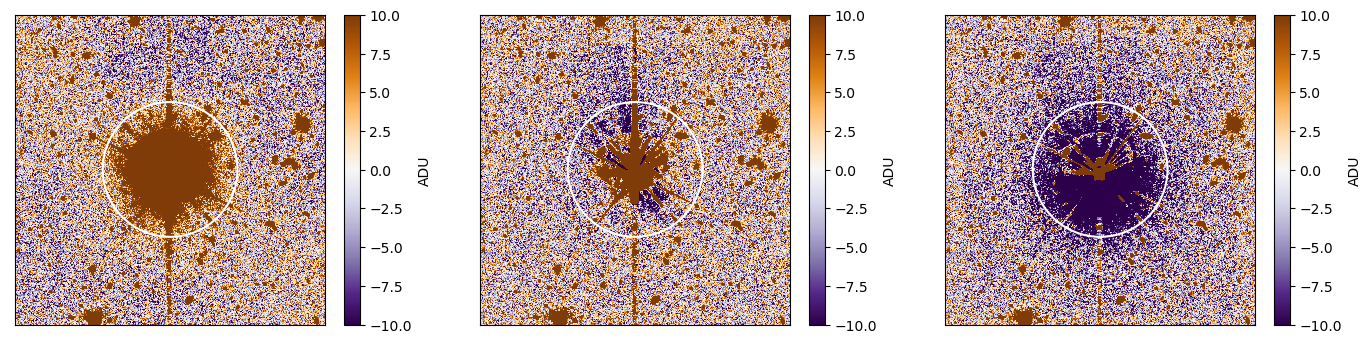}
   \end{center}
   \caption[example] 
   { \label{fig:f1_2} 
The left panel shows a star in an HSC image from a detector positioned near the edge of the focal plane (detector 038).
The middle and right panels display the same image after scaling and subtracting an extended PSF model.
The middle panel uses a detector-level model, while the right panel employs a full focal-plane-level model.
An identical white circle with a diameter of 250 pixels is added to the three panels for helping with comparison.
The oversubtraction in the lower region of the star in the right panel indicates that using a focal-plane-level model for a star from a detector near the edge of the focal plane can lead to inaccuracies, resulting in negative pixel values.
Conversely, using a model which is sensitive to the position of the bright star on the focal plane has significant benefits, as evidenced by the accuracy of the fit in the middle panel.
}
   \end{figure}

The left panel in Figure \ref{fig:f1_2} shows an HSC observation of a star (linearly scales), while the middle and right panels illustrate the results of subtracting extended PSF models from the same star.
The model used in the middle panel is a detector-level extended PSF, generated using stamps from images taken by the same detector as the star's observation.
This approach provides a localized model that accounts for characteristics specific to each detector.

Conversely, the right panel in Figure \ref{fig:f1_2} uses a full focal-plane-level extended PSF model derived from stamps of stars across all detectors on the focal plane.
As this model represents an average across the entire focal plane, it has a more generalized circular symmetry.
However, because the observed star in this particular detector is from a detector near the edge of the focal plane, its light distribution is inherently asymmetric.
Subtraction of a full focal-plane-level model results in oversubtraction in the lower region of the star here, as indicated by negative pixel values (purple colors).
This illustrates the risk of using generalized models for stars from detectors near the edge of focal plane for wide field observations.
Conversely, this approach highlights the strength in producing localized extended PSF models specific to particular positions across the focal plane.

These results, which are based on visual comparison, underscore the importance of choosing the appropriate extended PSF model when using the \texttt{SubtractBrightStarsTask}.
Detector-level models are often more accurate for star subtraction, especially for detectors near the edges of the focal plane where the model is asymmetric.
Conversely, full focal-plane-level models, while offering broader applicability and higher signal-to-noise ratios, may not capture localized asymmetries, leading to oversubtraction of stars near to the edge of focal plane.

To avoid such issues, it is crucial to consider the detector's position on the focal plane when selecting the appropriate extended PSF model.
This strategy helps ensure accurate star subtraction and prevents oversubtraction or other artifacts in the resulting calibrated exposures. 

\section{Results}
\label{sec:results}

The Bright Star Subtraction (BSS) pipeline was tested using Hyper Suprime-Cam (HSC) data to assess its effectiveness in modeling and subtracting extended Point Spread Functions (PSF) from bright stars.
The results reveal several key findings that inform plans for further optimization of the pipeline and its future development.

First, the extended PSF models show a significant dependency on the location of the detector on the camera's focal plane. Detectors near the center tend to produce more circularly symmetric extended PSF models, while those closer to the edges exhibit greater asymmetry.
This pattern indicates that a single, focal-plane-level extended PSF model is likely not to be suitable for all detectors, leading to issues like oversubtraction in some areas when utilized.

The choice of normalization annulus also plays a crucial role in modeling the extended PSF.
Smaller annuli can exclude stars due to saturation, while larger annuli may lower data quality because of reduced signal-to-noise ratios.
Our findings suggest that determining the optimal annulus size is complex but essential for the success of the phase 1 BSS pipeline.
Alternative methods for normalization may be worth considering in order to best address this challenge.

The use of detector-level extended PSF models yields more accurate results as compared to focal-plane-level models, especially in regions where detectors have distinct asymmetries.
However, the reduced number of stars available for modeling at the detector level relative to full focal-plane-level configurations can impact robustness.
This limitation indicates a need for further exploration of algorithms to improve the star count while maintaining localized accuracy.

In summary, the BSS pipeline shows promise in mitigating light contamination from bright stars in deep astronomical images.
However, the challenges of choosing appropriate normalization parameters and managing the reduced star count in detector-level PSF generation must be addressed to ensure high-quality results.
Future work should focus on further optimizing the normalization process and exploring methods to increase the robustness of detector-level PSF modeling.

\acknowledgments % equivalent to \section*{ACKNOWLEDGMENTS}       
 
This material is based upon work supported in part by the National Science Foundation through Cooperative Agreement AST-1258333 and Cooperative Support Agreement AST-1202910 managed by the Association of Universities for Research in Astronomy (AURA), and the Department of Energy under Contract No. DE-AC02-76SF00515 with the SLAC National Accelerator Laboratory managed by Stanford University. Additional Rubin Observatory funding comes from private donations, grants to universities, and in-kind support from LSSTC Institutional Members.

This work has made use of data from the European Space Agency (ESA) mission
{\it Gaia} (\url{https://www.cosmos.esa.int/gaia}), processed by the {\it Gaia}
Data Processing and Analysis Consortium (DPAC,
\url{https://www.cosmos.esa.int/web/gaia/dpac/consortium}). Funding for the DPAC
has been provided by national institutions, in particular the institutions
participating in the {\it Gaia} Multilateral Agreement.

% References
\bibliography{main} % bibliography data in report.bib
\bibliographystyle{spiebib} % makes bibtex use spiebib.bst

\end{document}